\documentclass[conference]{IEEEtran}

\usepackage{amsmath}
\usepackage{mathtools}
\usepackage{braket} 
\usepackage{cite}
\usepackage{algorithm}
\usepackage{algorithmic}

\usepackage{amsmath}   
\usepackage{amssymb}   
\usepackage{amsfonts}  

\usepackage{array}
\newcolumntype{P}[1]{>{\centering\arraybackslash}p{#1}}
\newcolumntype{M}[1]{>{\centering\arraybackslash}m{#1}}
\usepackage{subcaption}

\usepackage[table]{xcolor}
\usepackage[bookmarks=false]{hyperref}
\usepackage[top=0.75in, bottom=1in, left=0.625in, right=0.625in]{geometry}

\begin{document}

\title{Hybrid Quantum-Classical Maximum-Likelihood Detection via Grover-based Adaptive Search for RIS-assisted Broadband Wireless Systems}

\author{
    \IEEEauthorblockN{Maryam~Tariq\textsuperscript{\dag}, Raneem~Abdelrahim\textsuperscript{\dag}, Omar~Alhussein\textsuperscript{\ddag}, Sami~Muhaidat\textsuperscript{\dag,\S}}
    \IEEEauthorblockA{\textsuperscript{\dag}KU 6G Research Centre, Dept. of Computer and Information Engineering, Khalifa University, Abu Dhabi, UAE}
    \IEEEauthorblockA{\textsuperscript{\ddag}KU 6G Research Centre, Dept. of Computer Science, Khalifa University, Abu Dhabi, UAE\\
    $\textsuperscript{\S}$Department of Systems and Computer Engineering, Carleton University, Ottawa, Canada\\
    Emails: 100036622@ku.ac.ae, 100067007@ku.ac.ae, omar.alhussein@ku.ac.ae, sami.muhaidat@ku.ac.ae}
   
}







\maketitle

\begin{abstract}
The escalating complexity and stringent performance demands of sixth-generation wireless systems necessitate advanced signal processing methods capable of simultaneously achieving high spectral efficiency and low computational complexity, especially under frequency-selective propagation conditions. In this paper, we propose a hybrid quantum-classical detection framework for broadband systems enhanced by reconfigurable intelligent surfaces (RISs). We address the maximum likelihood detection (MLD) problem for RIS-aided broadband wireless communications by formulating it as a quadratic unconstrained binary optimization problem, that is then solved using Grover adaptive search (GAS). To accelerate convergence, we initialize the GAS algorithm with a threshold based on a classical minimum mean-squared error detector.
The simulation results show that the proposed hybrid classical-quantum detection scheme achieves \textit{near-optimal} MLD performance while substantially reducing query complexity. These findings highlight the potential of quantum-enhanced detection strategies combined with RIS technology, offering 
efficient and near-optimal solutions for broadband wireless communications.
\end{abstract}

\vspace{0.25em} 

\begin{IEEEkeywords}
broadband wireless, maximum likelihood detection (MLD), quadratic unconstrained binary optimization (QUBO), quantum computing
\end{IEEEkeywords}

\section{Introduction}
Driven by the ever-growing demand for higher data rates, wider coverage, and improved reliability, modern wireless systems face increasing pressure to overcome fundamental physical constraints such as limited bandwidth, low signal-to-noise ratio (SNR), and adverse propagation conditions. In wideband environments, frequency-selective fading induces severe inter-symbol interference (ISI), which poses a significant challenge to reliable communications. Cyclic-prefixed single-carrier (CPSC) transmission has emerged as a promising low-complexity alternative to orthogonal frequency division multiplexing, offering better performance in dispersive channels~\cite{falconer2002frequency, tubbax2001ofdm}. Meanwhile, reconfigurable intelligent surfaces (RISs) enable passive control of the wireless channel, thereby enhancing link quality without additional transmit power or bandwidth \cite{Jung2021On}.
As systems approach the practical limits of spectral and energy efficiency, the combination of CPSC and RIS offers a promising approach for high-performance communications in complex propagation environments. Although RIS phase configurations can strengthen specific propagation paths, they do not eliminate the frequency selectivity of the channel. RIS-aided CPSC systems still suffer from ISI, which necessitates efficient detection approaches to maintain adequate performance.

Classical detection strategies span a broad complexity-performance trade-off spectrum. Maximum likelihood detection (MLD) achieves optimal performance but incurs exponential complexity, scaling as \( \mathcal{O}(M^N) \) for block length \(N\) and constellation size \(M\), due to the exhaustive search over all symbol combinations \cite{1269977}. For channels with finite memory, the Viterbi algorithm offers an efficient implementation of MLD via trellis-based decoding, with complexity on the order of \( \mathcal{O}(NM^L) \), where \( L \) denotes the channel memory ~\cite{viterbi_comp}. Sphere decoding provides a complexity-aware alternative to MLD, but its performance depends heavily on channel conditions and may still be computationally intensive. Linear detectors such as zero-forcing (ZF) and minimum mean square error (MMSE) offer lower complexity, typically \( \mathcal{O}(N \log N) \) via fast-Fourier transform (FFT)-based equalization, but they are prone to noise amplification or residual ISI in highly dispersive channels~\cite{6259859}. These limitations motivate the exploration of advanced detection schemes that can balance optimality and complexity in realistic propagation environments.
 
Among emerging paradigms, quantum computing has recently gained attention as a promising direction for addressing complex optimization problems in wireless communications~\cite{10032075,9915359}. Particularly, quantum search algorithms, particularly Grover’s algorithm \cite{grover1996fast}, enable quadratic speedup for unstructured search problems via iterative amplitude amplification. Grover’s method operates by amplifying the probability amplitude of valid solutions through iterative oracle-based querying, allowing the complexity of exhaustive search to be reduced to \( \mathcal{O}(\sqrt{M^N}) \). Grover’s algorithm was later extended by Gilliam \emph{et al.} through Grover Adaptive Search (GAS) \cite{bulger2003implementing}, a framework designed to solve binary optimization problems using amplitude amplification guided by a cost function oracle. GAS enables Grover-style search over structured solution spaces by formulating the objective as a quadratic unconstrained binary optimization (QUBO) problem. To improve scalability and reduce quantum resource demands, Gilliam \emph{et al.} proposed a quantum dictionary-based representation for encoding polynomial cost functions with integer coefficients \cite{gilliam2020optimizing}. This enhancement facilitates efficient oracle construction for a broad class of optimization tasks, including QUBO and higher-order extensions, and improves the feasibility of GAS for near-term quantum hardware implementations.

This theoretical advantage makes it a promising tool for accelerating maximum likelihood detection, especially in high-dimensional search spaces arising from large modulation orders or block lengths. 
An early breakthrough in quantum-assisted wireless detection was introduced by Botsinis~\emph{et al.},  where quantum algorithms were employed to alleviate the computational complexity of MLD \cite{botsinis2013quantum}. Specifically, Grover-inspired search techniques, such as the Boyer–Brassard–Høyer–Tapp (BBHT) \cite{boyer1998tight}, and Dürr–Høyer (DH) algorithms \cite{durr1996quantum}, were used to perform MLD on quantum hardware, laying the foundation for subsequent research. In a related work, Botsinis~\emph{et al.} proposed a fixed-complexity quantum multi-user detection method for Code-Division Multiple Access (CDMA) and Space-Division Multiple Access (SDMA) systems, leveraging the DH algorithm and initializing the search with MMSE-based thresholds to reduce the number of quantum iterations \cite{6730885}. More recently, Norimoto~\emph{et al.} extended GAS to higher-order unconstrained binary optimization (HUBO), applying it to Multiple-Input and Multiple-Output (MIMO) MLD under gray-coded modulation, with detailed circuit-level analysis \cite{10044091}. In another study, Norimoto and Ishikawa addressed joint MLD in power-domain Non-Orthogonal Multiple Access (NOMA) systems by formulating the problem as a binary optimization task and improving GAS performance through a threshold design informed by the solution space distribution~\cite{10200430}.

While prior studies have demonstrated the feasibility of quantum-based or quantum-assisted MLD across various system models, they have not investigated key characteristics, such as multipath propagation, frequency selectivity, or RIS-aided architectures. To the best of our knowledge, this is the first work to consider frequency-selective propagation environment with RIS support.
%
%
Building on this, we propose a novel hybrid quantum-classical detection framework for RIS-assisted CPSC wireless systems. We first cast the considered problem as QUBO to enable efficient quantum handling via GAS. Furthermore, inspired by existing relevant literature, we leverage the low complexity of the MMSE detector in the classical domain to initialize a search threshold for GAS, effectively reducing the number of candidate solutions explored. Through quantum simulations, we show that the proposed approach achieves near-optimal MLD performance in terms of bit error rate, which significantly outperforms both the vanilla GAS algorithm (without proper threshold initialization) and the classical MMSE detector. This is achieved with substantially reduced query complexity compared to conventional MLD.

The remainder of this paper is organized as follows. Section \ref{sec:Background} provides essential background on quantum computing concepts, including qubits, quantum gates, and Grover’s search algorithm. Section \ref{sec:system_model} describes the system model for RIS-assisted  transmission. In Section \ref{Implementation}, the MLD problem is formulated as a QUBO and GAS is applied with real-valued encoding and MMSE-based initialization. Section \ref{Results} presents simulation results by comparing the proposed hybrid detector to classical baselines. Finally, Section \ref{conclusion} concludes the paper and outlines potential directions for future work.

\section{Quantum Preliminaries}
\label{sec:Background}
This section summarizes the quantum concepts required for the proposed
detection framework: qubits and superposition, basic quantum gates, and
Grover’s search algorithm.  

\subsection{Qubits and Superposition}

A quantum bit (qubit) is a two‑level quantum system whose state
is described by a unit vector in Hilbert space
$\mathcal{H}$.  
A single‑qubit state is written as
\begin{equation}
  \ket{\phi} = \alpha\ket{0} + \beta\ket{1}, \quad
  \alpha,\beta\in\mathbb{C},\;
  |\alpha|^{2}+|\beta|^{2}=1.
\end{equation}
For an $n$‑qubit register, the composite state resides in the tensor
product space $\mathcal{H}^{\otimes n}$, enabling
\emph{superposition} which is a key resource for quantum parallelism.
A measurement in the computational basis collapses $\ket{\phi}$ to
$\ket{0}$ with probability $|\alpha|^{2}$ or to $\ket{1}$ with
probability $|\beta|^{2}$.

\subsection{Quantum Gates and Circuits}
Quantum algorithms are realized as unitary circuits acting on qubit
registers. The gates used in this paper are

\noindent \textit{Hadamard Gate, $H$}: Given $\ket{0}$ or $\ket{1}$, it creates a state in superposition as follows:
\begin{equation}\label{eq:hadamard}
H\ket{0} = \frac{\ket{0}+\ket{1}}{\sqrt{2}}, \,
H\ket{1} = \frac{\ket{0}-\ket{1}}{\sqrt{2}}.
\end{equation}
\noindent\textit{Pauli-$X$ Gate, $X$}: Often called the quantum \textsc{NOT}, it swaps the computational basis states:
\begin{equation}\label{eq:pauliX}
X\ket{0} = \ket{1}, \quad X\ket{1} = \ket{0}.
\end{equation}

\noindent\textit{Pauli-$Y$ Gate, $Y$}: Combines a bit flip with a $\tfrac{\pi}{2}$ phase shift, sending $\ket{0}\!\mapsto\! i\ket{1}$ and $\ket{1}\!\mapsto\! -i\ket{0}$. The Pauli-$Z$ Gate $Z$ leaves $\ket{0}$ unchanged while adding a $\pi$ phase to $\ket{1}$, i.e., $\ket{1}\!\mapsto\! -\ket{1}$.

\noindent\textit{Controlled‑NOT Gate (CNOT)}: flips a target qubit conditional on a control qubit being $\ket{1}$.

\noindent \textit{Toffoli Gate (CCX)}: a universal three‑qubit gate that flips the target when both control qubits are $\ket{1}$.

These gates form a universal set for quantum computation and are used to
implement the QUBO-based oracle and diffusion operators in Grover’s algorithm.

\subsection{Grover Search}
Grover’s search algorithm finds a marked item in an unsorted database of
size $2^{n}$ using $\mathcal{O}(2^{n/2})$ oracle calls, which is quadratically
faster than the classical $\mathcal{O}(2^{n})$ bound.
Let $\mathcal{X}=\{0,\dots,2^{n}-1\}$ and
$f:\mathcal{X}\!\to\!\{0,1\}$ be a Boolean oracle such that
$f(x^{\star})=1$ for the unique solution $x^{\star}$ and $0$ otherwise. Here, the oracle can be simply expressed as
\begin{equation}
  O_{f}\ket{x} = (-1)^{f(x)}\ket{x},\quad x\in\mathcal{X}.
\end{equation}

Grover search works as follows: First, prepare an equal superposition of all states through tensor product of Hadamard gates, i.e.,
\begin{equation}
  \ket{\psi_{0}}
   = H^{\otimes n}\ket{0}^{\otimes n}
   = 2^{-n/2}\sum_{x\in\mathcal{X}}\ket{x}.
\end{equation}
Then, we apply a Grover iteration that consists of two gates, namely the oracle gate and the diffusion gate. One Grover iteration applies
\begin{equation}
  G = D\,O_{f},
\end{equation}
where $O_{f}$ marks $\ket{x^{\star}}$ with a phase $-1$ and
$D = 2\ket{\psi_{0}}\!\bra{\psi_{0}} - I_{2^{n}}$  
is the diffusion operator which effectively inverts the amplitude of each state about its mean. Effectively, since the diffusion operator inverts all values about their mean, the marked state gets amplified, whereas the amplitudes of other states reduce.

\textit{Query Complexity} --
After $k^{\star} =  \frac{\pi}{4}\,2^{n/2}$
iterations, the probability of measuring $x^{\star}$ exceeds~0.5. Thus, Grover search requires $\mathcal{O}(2^{n/2})$ oracle queries, offering a quadratic speed‑up over exhaustive search.

\section{System Model and Problem Description}
\label{sec:system_model}

A frequency-selective communication system is considered, where the base station (BS) and user equipment (UE) are each equipped with a single antenna. Communication is established exclusively via a RIS comprising \( R \) passive reflecting elements. The direct BS-UE link is assumed to be unavailable, and all signal propagation occurs through RIS-assisted paths.
The discrete-time channel impulse responses (CIRs) between the BS and the \( r^{\text{th}} \) RIS element, and between the RIS element and the UE, are denoted by \( \mathbf{h}_{\text{BI}}^{(r)} \in \mathbb{C}^{L_{\text{BI}} \times 1} \) and \( \mathbf{h}_{\text{IU}}^{(r)} \in \mathbb{C}^{L_{\text{IU}} \times 1} \), respectively, where $L_{\text{BI}}$ and $L_{\text{IU}}$ denote the lengths of the respective CIRs. The overall cascaded channel is formed by convolving the individual CIRs and applying the RIS phase shifts, yielding the effective end-to-end channel:
\begin{equation}
\mathbf{\tilde{h}}_{\text{}} = \sum_{r=1}^{R} \left( \mathbf{h}_{\text{BI}}^{(r)} * \mathbf{h}_{\text{IU}}^{(r)} \right) \phi_r \in \mathbb{C}^{L_{\text{}} \times 1},
\end{equation}
where `$*$' denotes the discrete-time convolution operation, 
$
L = L_{\text{BI}} + L_{\text{IU}} - 1
$ represents the effective number of channel taps, and \( \phi_r = e^{j\theta_r} \) is the complex reflection coefficient applied by the \( r^{\text{th}} \) RIS element. The phase \( \theta_r \in [0, 2\pi) \) is chosen to compensate for the phase of the first tap of each cascaded path, enabling constructive combining at the receiver. The BS maps $b = N\log_2(M)$ information bits into an $M$-ary modulated vector $\mathbf{x}=[x_1,\dots,x_N]^T \in \mathbb{C}^{N\times 1}$, where each symbol $x_n \in \mathcal{X}$ is drawn from an $M$-ary constellation. A cyclic prefix (CP) of length \( L_{\text{cp}} \geq L \) is prepended to \( \mathbf{x} \), and the condition \( N \geq L \) is assumed to ensure the validity of the circular convolution model. The CP-augmented signal is denoted by \( \mathbf{x}_{\text{cp}} \in \mathbb{C}^{(N+L_{\text{cp}})\times1} \). After perfect CP removal at the receiver, the channel is represented as a circulant matrix \( \tilde{\mathbf{H}} \in \mathbb{C}^{N\times N} \), and the received signal is given by

\begin{equation}
\mathbf{y} = \mathbf{\tilde{H}} \mathbf{x} + \mathbf{w},
\end{equation}
where \( \mathbf{w} \sim \mathcal{CN}(0, \sigma_w^2 \mathbf{I}) \) is additive white Gaussian noise. Given the received signal model in (8), detection is performed using the optimal yet computationally demanding MLD, which serves as a classical performance benchmark. The optimal transmit vector is obtained by solving the MLD problem:
\begin{equation}
\hat{\mathbf{x}} = \arg \min_{\mathbf{x} \in \mathcal{X}^N} \left\| \mathbf{y} - \tilde{\mathbf{H}} \mathbf{x} \right\|^2,
\label{eq:D}
\end{equation}
where \( \mathcal{X} \) denotes the modulation alphabet. This formulation entails an exhaustive search over all \( M^N \) possible symbol combinations. As a result, systems with larger delay spreads require longer block lengths, thereby increasing the complexity of MLD. Consequently, MLD is rendered impractical for highly frequency-selective channels or high-throughput transmission scenarios.


\section{MLD via Grover-based Adaptive Search}
\label{Implementation}

\subsection{Casting MLD as QUBO}
To convert the problem into a form suitable for quantum optimization, we reformulate it as a QUBO problem.
For a system characterized by the received signal vector \( \mathbf{y} \in \mathbb{C}^N \), channel matrix \( \mathbf{\tilde{H}} \in \mathbb{C}^{N \times N} \), and transmitted symbol vector \( \mathbf{x} \in \{-1, +1\}^N \), the MLD is formulated as the problem of minimizing the squared Euclidean distance as defined in \eqref{eq:D}, expanding the squared norm yields
\begin{subequations}\label{eq:euclidean}
\begin{align}
\|\mathbf{y}-\tilde{\mathbf{H}}\mathbf{x}\|^2
&= (\mathbf{y}-\tilde{\mathbf{H}}\mathbf{x})^\dagger(\mathbf{y}-\tilde{\mathbf{H}}\mathbf{x}) \label{eq:euclidean_a}\\
&= \mathbf{y}^\dagger\mathbf{y} - 2\Re\{\mathbf{x}^\dagger \tilde{\mathbf{H}}^\dagger \mathbf{y}\} + \mathbf{x}^\dagger \tilde{\mathbf{H}}^\dagger \tilde{\mathbf{H}} \mathbf{x}. \label{eq:euclidean_b}
\end{align}
\end{subequations}
Since the term \( \mathbf{y}^\dagger\mathbf{y} \) is constant with respect to \( \mathbf{x} \), it can be omitted from the optimization. The objective function becomes
\begin{equation}
E(\mathbf{x}) = \mathbf{x}^\dagger \tilde{\mathbf{H}}^\dagger \tilde{\mathbf{H}} \mathbf{x} - 2\Re\{\mathbf{x}^\dagger \tilde{\mathbf{H}}^\dagger \mathbf{y}\}.
\end{equation}
We define the matrix and vector terms,
\begin{equation}
\mathbf{Q}_{\text{BPSK}} = \tilde{\mathbf{H}}^\dagger\tilde{\mathbf{H}}, \quad \mathbf{c}_{\text{BPSK}} = -2\Re\{\tilde{{\mathbf{H}}}^\dagger\mathbf{y}\},
\end{equation}
leading to the quadratic optimization form:
\begin{equation}
\min_{\mathbf{x} \in \{-1, +1\}^N} \mathbf{x}^\dagger \mathbf{Q}_{\text{BPSK}} \mathbf{x} + \mathbf{c}_{\text{BPSK}}^\top \mathbf{x}.
\end{equation}

For modulation schemes such as BPSK and QPSK, each transmitted symbol is linearly mapped from binary bits. This ensures the cost function retains a quadratic structure, as the signal model involves a linear transformation followed by a squared Frobenius norm.
Since quantum computers operate on binary variables \( \mathbf{b} \in \{0, 1\}^N \), we substitute \( \mathbf{x} = 2\mathbf{b} - 1 \). Applying this transformation yields
\begin{equation}
\min_{\mathbf{b} \in \{0, 1\}^N} (2\mathbf{b} - 1)^\dagger \mathbf{Q}_{\text{BPSK}} (2\mathbf{b} - 1) + (2\mathbf{b} - 1)^\top \mathbf{c}_{\text{BPSK}}.
\end{equation}
Expanding the terms leads to the cost function formulation,
\begin{equation}
E(\mathbf{b}) = \mathbf{b}^\top Q' \mathbf{b} + \mathbf{c'}^\top \mathbf{b} + \text{const},
\end{equation}
where \( Q' \) and \( \mathbf{c'} \) are appropriately redefined matrices and vectors absorbing the constants from the transformation. This binary quadratic form is well suited for quantum-enhanced optimization algorithms such as GAS.

To implement this QUBO problem on gate-based quantum hardware, the binary variables \( b_i \) are encoded using Pauli-Z operators through the transformation\footnote{The detailed circuit construction for mapping the QUBO formulation into a Grover oracle is omitted here due to space constraints and will be presented in an extended journal version.} \( b_i = (1 - Z_i)/2 \). Substituting this into the QUBO cost function results in the following Ising Hamiltonian:
\begin{equation}
\mathcal{H}
 = \sum_i h_i Z_i + \sum_{i<j} J_{ij} Z_i Z_j + \text{const},
\end{equation}
where \( h_i \) and \( J_{ij} \) are real-valued coefficients derived from the QUBO formulation. This Hamiltonian form is native to many quantum algorithms and hardware platforms.

\subsection{Grover Adaptive Search}
GAS \cite{Gilliam2021groveradaptive} is an iterative quantum optimization framework designed to minimize discrete objective functions. These functions take the form \( E(b): \mathbb{B}^n \rightarrow \mathbb{Z} \), where \( b \in \mathbb{B}^n \) is a binary vector consisting of \( n \) bits, and \( E(b) \) is a cost function that assigns an integer value to each possible binary configuration. Lower values of  \( E(b) \) typically correspond to better solutions, reflects the quality of the candidate \( b \).

In the standard GAS formulation, \( n \) qubits are used to encode the binary decision variables, and an additional \( m \) qubits are allocated to represent the cost function output using two’s complement encoding, which supports both positive and negative integers.
At each iteration, GAS initializes the quantum state into a uniform superposition across all binary inputs, while simultaneously encoding the shifted cost function $E(b) - y_i$ in an ancillary register:
\begin{equation}
    A_{y_i} \ket{0}^{\otimes n} \ket{0}^{\otimes m} = \frac{1}{\sqrt{2^n}} \sum_{b=0}^{2^n - 1} \ket{b}_n \ket{E(b) - y_i}_m,
\end{equation}
where $y_i \in \mathbb{Z}$ is a dynamically updated threshold corresponding to the best function value observed so far. The operator $A_{y_i}$ consists of Hadamard gates applied to the quantum registers, followed by a value-encoding circuit constructed using controlled-phase rotations, and the inverse quantum Fourier transform (IQFT). The value-encoding circuit  maps the cost function \( E(b) \) into quantum phases using a structured composition of controlled-phase rotation gates defined over an \( m \)-qubit register, where \( m \) determines the precision of the encoded value. Each term in the polynomial objective, whether constant, linear, or higher-order, is implemented via a composition of single- and multi-controlled phase gates with rotation angles determined by the corresponding integer coefficients.

\begin{figure}[]
    
    \includegraphics[width=1\linewidth]{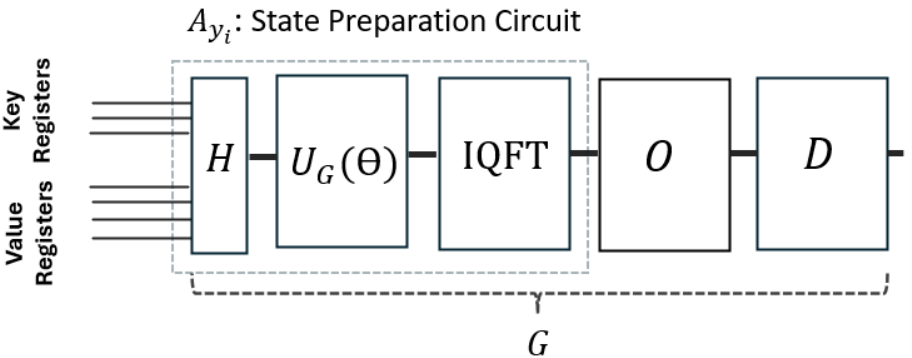}
    \caption{Grover adaptive search circuit showing state preparation $A_{y_i}$, oracle $O$, and diffusion operator $D$, iterated to amplify low-cost solutions based on threshold $y_i$.}
    
    \label{fig:grover}
\end{figure}

For a constant coefficient \( k \in \mathbb{Z} \), the corresponding quantum phase is given by
\begin{equation}
    \theta = \frac{2\pi k}{2^m}.
    \label{eq:Integarencoding}
\end{equation} 
Operator \( U_G(\theta) \) applies this phase encoding as
\begin{equation}
    U_G(\theta) = R(2^{m-1}\theta) \otimes R(2^{m-2}\theta) \otimes \cdots \otimes R(\theta),
\end{equation}
where \( R(\theta) \) is the basic phase rotation gate.
%
The oracle component of the Grover operator identifies basis states for which $E(b) < y_i$ by leveraging the sign bit of the two’s complement representation. Specifically, a single-qubit phase flip (Z-gate) is applied conditioned on this sign qubit being in the $\ket{1}$ state. The full Grover iteration is then constructed as:
\begin{equation}
    G = A_{y_i}  D  A_{y_i}^\dagger  O,
\end{equation}
where $D$ denotes the standard diffusion operator and $O$ is the oracle described above. The number of Grover iterations $L_i$ at each round is sampled uniformly from $\{0, 1, \ldots, \lceil k - 1 \rceil\}$, with $k$ being an adaptive control parameter. The complete quantum circuit for a single Grover iteration, including cost encoding, inverse QFT, and amplitude amplification, is illustrated in Fig.~\ref{fig:grover}

The GAS procedure updates the threshold $y_i$ when a better solution is found. If no improvement occurs, $k$ is scaled by a factor $\lambda > 1$ (typically $\lambda = 8/7$), gradually increasing the search scope. 
The full algorithm is summarized in Algorithm~\ref{alg:real_valued_gas}.
\begin{algorithm}[t]
\caption{ Grover Adaptive Search \cite{Gilliam2021groveradaptive} }
\label{alg:real_valued_gas}
\begin{algorithmic}[1]
\REQUIRE $E : \mathbb{B}^n \rightarrow \mathbb{R}$, $\lambda = 8/7$
\ENSURE $\mathbf{b}$
\STATE Uniformly sample $\mathbf{b}_0 \in \mathbb{B}^n$ and set $y_0 = E(\mathbf{b}_0)$ \hfill 
\STATE Set $k = 1$ and $i = 0$
\REPEAT
    \STATE Randomly select the rotation count $L_i$ from the set $\{0, 1, \ldots, \lceil k - 1 \rceil \}$
    \STATE Evaluate $\mathbf{G}^{L_i} A_{y_i} |0\rangle_{n+m}$ and obtain $\mathbf{b}$
     \hfill 
    \IF{$y < y_i$}
        \STATE $\mathbf{b}_{i+1} = \mathbf{b}$, $y_{i+1} = y$, and $k = 1$
    \ELSE
        \STATE $\mathbf{b}_{i+1} = \mathbf{b}_i$, $y_{i+1} = y_i$, and $k = \min\{\lambda k, \sqrt{2^n}\}$
    \ENDIF
    \STATE $i = i + 1$
\UNTIL a termination condition is met
\end{algorithmic}
\end{algorithm}

\vspace{0.5em}
\noindent
The quantum advantage of GAS lies in its ability to amplify promising solution states through iterative amplitude amplification, targeting those with objective values below the current threshold. Proper selection of the number of function-evaluation qubits $m$ is critical to ensuring correct arithmetic over the range of values $E(b) - y_i$, which can span from $E_{\text{min}} - E_{\text{max}}$ to $E_{\text{max}} - E_{\text{min}}$. The relation given by inequality 
\begin{equation}
    -2^{m-1} \leq E(b) - y_i < 2^{m-1}
    \label{eq:Value}
\end{equation}
must hold to avoid overflow during encoding, ensuring the validity of the two’s complement representation.

In many practical optimization problems, particularly in wireless communications and signal detection, the objective function $E(b)$ comprises real-valued coefficients due to underlying physical models involving continuous-valued channel matrices, noise statistics, or signal representations. Our case is no different; the cost function in MLD includes Euclidean distances derived from complex-valued channel outputs, which yield inherently real-valued quantities.
To extend GAS to handle such real-valued functions, two encoding strategies have been proposed by Gilliam \textit{et al.}, namely (\textit{i}) integer approximation and (\textit{ii}) direct real-valued encoding \cite{Gilliam2021groveradaptive}.

\paragraph{Integer Approximation}
In this approach, real-valued coefficients are first approximated as rational numbers and then scaled by a common denominator to yield integer coefficients. While mathematically sound, this method increases the dynamic range of the cost function, thereby requiring a larger number of value qubits $m$ to avoid overflow.
As $m$ grows, so does the quantum circuit depth and the number of multi-controlled phase gates, limiting scalability and making implementation more susceptible to noise.

\paragraph{Direct Encoding}
To retain numerical fidelity while minimizing qubit requirements, a direct encoding method maps each real-valued coefficient $a \in \mathbb{R}$ to a quantum phase using
\begin{equation}
    \theta = \frac{2\pi a}{2^m},
    \label{eg:Realencoding}
\end{equation}
which is used in constructing the phase gate $U_G(\theta)$. After applying the inverse quantum Fourier transform (IQFT), the resulting state is not a single basis state as in the integer case, but a superposition characterized by the Fejér distribution \cite{Gilliam2021groveradaptive}:
\begin{equation}
    U_{\text{Fejér}}(\theta)\ket{0}^{\otimes m} = \sum_{l=0}^{2^m-1} \left\langle g(\theta), g\left(\frac{2\pi l}{2^m}\right) \right\rangle \ket{l},
\end{equation}
where $g(\theta) = \left[1, e^{j\theta}, \cdots, e^{j(2^m-1)\theta} \right]/\sqrt{2^m}$. This inner product leads to constructive interference near $\theta$ and destructive interference elsewhere, forming a smooth peaked distribution. Consequently, measurement of the $m$-qubit value register yields values clustered around the true $E(b)$.

Due to the inherent spread of the Fejér distribution, the quantum measurement may yield an estimate of $E(b)$ that is lower than its true value. This can prematurely reduce the GAS threshold $y_i$, potentially excluding all valid candidate solutions from further amplification. To mitigate this, following \cite{10044091}, the algorithm is modified to discard the quantum-evaluated cost $y$ and instead compute the exact value classically.
This extension broadens the scope of GAS well beyond integer QUBO, enabling its application to a wide class of problems requiring real-valued objective formulations.

\subsection{Adaptive Threshold Initialization}
We replace the conventional random threshold with a \emph{deterministic} one derived from a classical MMSE detector.  
The received block is first MMSE‑equalized, yielding a hard‑decision bit vector \( \bar{\mathbf{b}}_{0} \).  
Its QUBO cost is computed as $y_{0} = E\bigl(\bar{\mathbf{b}}_{0}\bigr)$,
%
which is then adopted as the initial GAS threshold.

Operating in the frequency domain, the MMSE filter is realized as a diagonal matrix \( \mathbf{\Phi} \in \mathbb{C}^{N \times N} \) with entries
\begin{equation}
\Phi(i, i) = \frac{\Lambda^*(i)}{|\Lambda(i)|^2 + \sigma_w^2},
\end{equation}
where \( \sigma_w^2 \) is the noise variance and \( \Lambda(i) \) denotes the \( i \)-th diagonal element of the frequency-domain channel matrix $\mathbf{\Lambda} = \mathbf{F}\mathbf{H}$, where $\mathbf{F}$ is the discrete fourier transform matrix.
The equalized signal is computed as \( \mathbf{d}_f = \mathbf{\Phi} \mathbf{y}_f \), and the final symbol estimates are obtained by applying the inverse FFT followed by hard decision decoding.

Although MMSE offers reduced complexity relative to MLD, it often yields symbol estimates close to the true solution.  
Using \( y_{0} \) as a threshold sets a tight initial upper bound on the QUBO objective, which limits the number of marked states and improves the efficiency of amplitude amplification.  
As a result, the expected number of Grover iterations required to locate the optimal solution is significantly reduced~\cite{6730885}.

\section{Simulation Results}\label{Results}
\begin{figure}[]
    \includegraphics[width=0.8\linewidth]{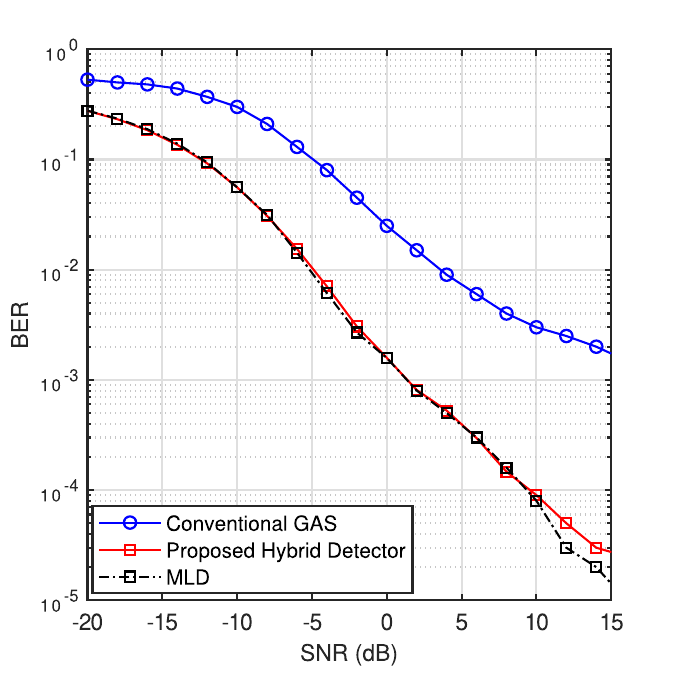}
    \caption{Comparison between the proposed hybrid detector, conventional GAS, and classical MLD for three RIS configurations: no RIS, \( R = 4 \), and \( R = 8 \), with \( N = 3 \) symbols.
    }
    \label{fig:grover_ml_framework}
    \vspace{-1em}
\end{figure}

This work was evaluated through simulations conducted on a quantum statevector simulator using IBM Qiskit. This noiseless backend provides exact quantum amplitudes, allowing us to isolate the algorithmic behavior of Grover-based search without the confounding effects of hardware noise or decoherence. The considered system model follows the RIS-assisted single-antenna communication setup described in Section III, and BPSK modulation is employed throughout the simulations. Given the considered system model, Fig.~\ref{fig:grover_ml_framework} compares the BER performance of the proposed hybrid quantum-classical detector against the conventional GAS and classical MLD across a broad range of SNRs. Conventional GAS exhibits a significantly higher BER for all SNR values, clearly highlighting its limitations in efficiently handling real-valued optimization problems inherent to frequency-selective communication scenarios. Conversely, the proposed hybrid detector closely approaches the optimal performance of classical MLD, particularly at moderate-to-high SNR levels (above 0 dB), where the performance gap becomes negligible. This result demonstrates the effectiveness of the proposed hybrid approach, reducing the computational burden while preserving near-optimal detection.

\begin{figure}[]
    \includegraphics[width=0.8\linewidth]{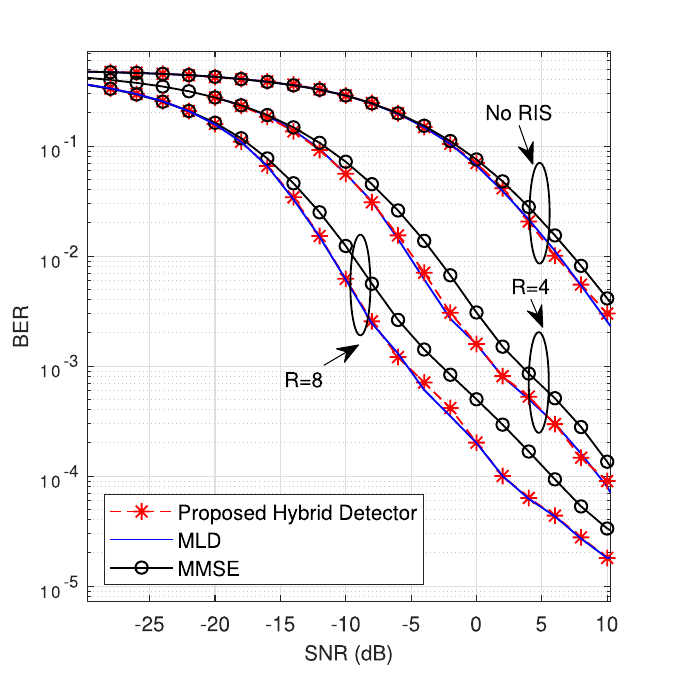}
    \caption{BER comparison of the proposed hybrid quantum-classical detector, MLD, and MMSE detector for varying RIS sizes (\( R = 0, 4, 8 \)) with BPSK and \( N = 3 \).}
    \label{fig:PropMLD_MMSE}
\end{figure}

Fig.~\ref{fig:PropMLD_MMSE} further illustrates the impact of incorporating an RIS and varying the number of RIS reflecting elements $R$ on system performance. The simulation results confirm significant performance enhancements obtained by employing RIS within the communication framework. As the RIS element count increases from $R=4$ to $R=8$, a pronounced BER reduction is achieved across all detectors. For example, at an SNR of $-5$ dB, increasing RIS elements from $4$ to $8$ reduces the BER from approximately $2\times10^{-3}$ to about $3\times10^{-4}$, clearly underscoring the substantial benefits offered by RIS in terms of diversity and signal enhancement. Additionally, the proposed hybrid detector consistently outperforms classical MMSE detection under all RIS configurations, confirming its effectiveness in leveraging quantum-enhanced optimization for RIS-based channels.

\section{Conclusion}
\label{conclusion}
We have developed a hybrid quantum-classical detection framework for RIS-assisted broadband wireless communications. By formulating the detection task as a QUBO problem and employing GAS with an MMSE-initialized threshold, the proposed method efficiently addresses the prohibitive computational complexity inherent in exhaustive MLD techniques. Our simulations validate the near-optimal detection capability and computational advantage of the hybrid approach, demonstrating its practical scalability to realistic broadband scenarios with large block lengths and higher-order modulations. Looking forward, further research can include extending this framework to more complex wireless scenarios, exploring robustness against quantum hardware imperfections, and validating the proposed architecture on emerging quantum computing platforms.

\bibliographystyle{IEEEtran}
\bibliography{References}

\end{document}